*Review*

# Occam's Razor for *Big Data*? On Detecting Quality in Large Unstructured Datasets

**Birgitta Dresp-Langley [1], Ole Kristian Ekseth [2,\*], Jan Fesl [3], Seiichi Gohshi [4], Marc Kurz [5] and Hans-Werner Sehring [6,\*]**

[1] Centre National de la Recherche Scientifique, UMR 7357 ICube Lab, CNRS-Strasbourg University, 67200 Strasbourg, France

[2] NTNU Trondheim, 7491 Trondheim, Norway

[3] Institute of Applied Informatics, Faculty of Science, University of South Bohemia Czech Republic, 370 05 České Budějovice, České

[4] Informatics Department, Kogakkan University, Ise Mie 516-0016, Japan

[5] Department of Mobility & Energy, University of Applied Sciences Upper Austria, 4232 Hagenberg Austria

[6] Namics—A Merkle Company, 20357 Hamburg, Germany

**\*** Correspondence: oekseth@gmail.com (O.K.E.); hans-werner.sehring@namics.com (H.-W.S.)

**Abstract:** Detecting quality in large unstructured datasets requires capacities far beyond the limits of human perception and communicability and, as a result, there is an emerging trend towards increasingly complex analytic solutions in data science to cope with this problem. This new trend towards analytic complexity represents a severe challenge for the principle of parsimony (*Occam's razor*) in science. This review article combines insight from various domains such as physics, computational science, data engineering, and cognitive science to review the specific properties of *big data*. Problems for detecting data quality without losing the principle of parsimony are then highlighted on the basis of specific examples. Computational building block approaches for data clustering can help to deal with large unstructured datasets in minimized computation time, and meaning can be extracted rapidly from large sets of unstructured image or video data parsimoniously through relatively simple unsupervised machine learning algorithms. Why we still massively lack in expertise for exploiting *big data* wisely to extract relevant information for specific tasks, recognize patterns and generate new information, or simply store and further process large amounts of sensor data is then reviewed, and examples illustrating why we need subjective views and pragmatic methods to analyze *big data* contents are brought forward. The review concludes on how cultural differences between East and West are likely to affect the course of *big data* analytics, and the development of increasingly autonomous artificial intelligence (AI) aimed at coping with the *big data* deluge in the near future.

## 1. Introduction

The Cisco Global Cloud Index 2016–2021 Forecast [1] estimates that nearly 850 zeta bytes (ZB) of data will be generated by all people, machines, and things by 2021, up from the 220 ZB generated in 2016. Most of the more than 850 ZB generated by 2021 will be "ephemeral in nature" and will be "neither saved nor stored". The Cisco forecast states further that "most of this ephemeral data is deemed not useful to save", and that "approximately 10 percent of it is useful, which means that there will be 10 times more useful data being created (85 ZB, 10 percent of the 850 total) than will be stored or used (7.2 ZB) in 2021". Apart from the fact that grasping the significance of this statement in itself represents a challenge, the problem of identifying what is and what is not useful in the growing jungle of information overflow represents, indeed, one of the most pressing challenges for science, business, and society. As a consequence, the *big data* issue, coupled with that of finding new data analytics, radically challenge established theory and practice across the sciences, engendering a new form of scientific uncertainty and paradigm shifts [2] in all major fields of science, from physics to the humanities. Already more than 10 years ago, Anderson [3], among other visionaries, predicted that the *big data* deluge will, ultimately, lead to the end of science and its quest for causality. Whole cohorts of freshly recruited data scientists will occupy their time looking for correlations in non-dimensional input, and any scientist trained the "old-fashioned way" knows only too well that correlation does not imply causality. Instead, universities and research labs now train the young to cope with the volume–variety–velocity–veracity–value (the 5-'v' problem) of a data-driven science and society (Figure 1), which the *big data* deluge has brought upon us without asking for our opinion.

Getting a grip on the many problems represented by *big data* involves being able to detect what is and what is not meaningful in a more and more complex jungle of facts and figures. The effective processing of an increasingly large amount of data, accumulating in the cloud and, in principle, designated to stay there, requires capacities far beyond the limits of human perception and communicability. *Big data* represent a number of problems of steadily increasing magnitude for society, which call for new analytic approaches and conceptual solutions, and science urgently needs to develop the necessary expertise, methods, and procedures to ensure that the data that will ultimately be retained for a useful exploitation will be of the best possible quality, convey genuine meaning, and produce beneficial effects on both science and society.

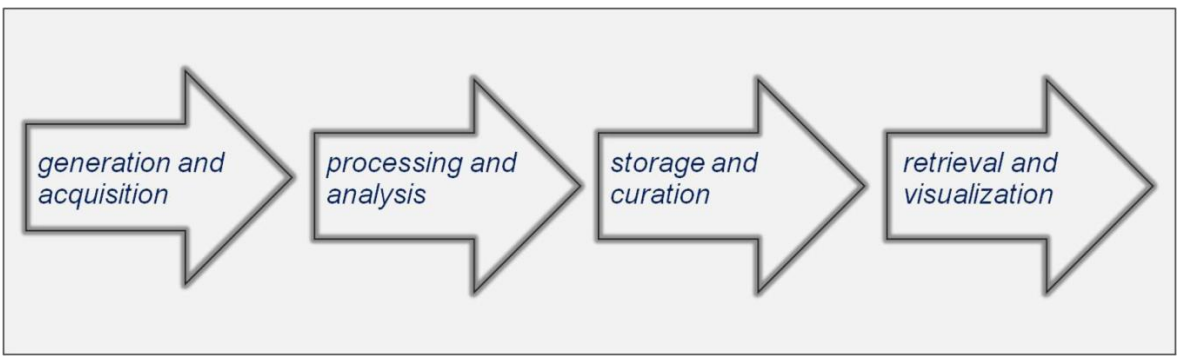

**Figure 1.** The *big data* value chain [1]. All steps in the chain, from data generation to retrieval and visualization for exploitation to the benefit of science, business, and society, are subject to problems relative to the *volume, variety, velocity,* and *veracity* of the data, affecting their ultimate *value* to science and society (5-'v' problem).

The principle of parsimony, or *Occam's razor*, stems from writings on the logic of explanation by the cleric William of Occam [4]. In science, *Occam's razor* reflects both a general rule of prudence and a conservative guideline for investigation that consists of aiming for the simplest among possible explanations for, or model approaches to, the phenomenon under study. While the principle of parsimony has been implicitly adhered to across and within all major scientific disciplines (mathematics, biology, physics), its fundamentals and application are severely challenged by the emerging trend towards increasingly complex analytic solutions in data science within the context of *big data*. This new problem space requires new strategies of data analysis and requires maintaining an open mind to alternative possibilities, a new philosophy as to how data analysis is be carried out, rather than any fixed set of method [5,6]. The complexity of the *big data* problem space is such, that there can be no single, novel or established, solution for data quality control [5], and there is an urgent need for wider critical reflection within the scientific community on the epistemological implications of this unfolding data revolution along with the rapid changes in research practice presently taking place [3].

This collaborative review article results from an international panel consortium effort at the 'Cognitive 2019' conference in Venice [7] and represents a "think-tank" approach towards critical conceptual knowledge construction on some of the problems represented by *big data*. The paper benefits from insight from various domains such as physics, computational science, data engineering, and cognitive neuroscience. Section 1 summarizes the specific properties of *big data* and the problems these pose for science and, in particular, applied data science. Section 2 provides a synthetic overview of big data analytic solutions from machine learning to artificial intelligence, followed by a more detailed discussion of two examples of parsimonious analytics in the *big data* context: Clustering for determining data structure of any kind, and parsimony-driven analysis of single-pixel change in large sets of image data with minimalistic artificial intelligence. Why we still massively lack in expertise for exploiting *big data* wisely in order to extract relevant information for specific tasks, recognize patterns and generate new information, or simply store and further process large amounts of sensor data is summarized in Section 3 on the example of the *smart city* concept. Issues relative to model building and reasons why we need subjective views and pragmatic methods to analyze *big data* contents are brought forward in Section 4. Section 7 concludes the review by evoking a few thoughts on how cultural differences between East and West are likely to affect the course of *big data* analytics, and the development of increasingly autonomous artificial intelligence (AI) in the near future. *Big data* represents a problem our cultural development has brought upon us, and without this problem, there would be no need for highly developed artificial intelligence.

## 2. Testing *Big Data* Quality: Potential and Limitations

The *big data* concept is a new trend that occurs in many scientific directions. It may seem that such a concept has been around for many years but, in reality, the *big data* of today have created radically new challenges and also pose a number of new problems. These are directly related to their properties, which are the following:

(a) *Uniqueness*—the volumes of *big data*sets are mainly presented in the scale of EXA, PETA, or ZETA bytes [8]. This means that such datasets are retrieved from many specific unique sources, which cannot be easily exchanged by other optional e. q. values from sensor networks located in a specific area. Another reason for is related to the time necessary for the data production and the collection time, and it is not rare that these move into the range of tens of months or more [9].

(b) *A-dimensionality*—*big data* mostly have no concrete structure, are unsorted, and their value distribution functions are typically unknown [10]. For many *big data* types, it is impossible to sort the data according to the value of a specific parameter (e.g., speech samples, pictures), because they are not straightforwardly comparable.

(c) *Specificity*—this feature has much in common with data uniqueness. In general, the datasets are retrieved from many sources, and their content is quite specific. This means that what is valid for one dataset need not be for another. This specificity can have many different reasons—like the data type, resource type, geographical context [11], or other.

(d) *Cost*—their storing and processing requires large and expensive, high-capacity data stores and powerful distributed computing systems.

(e) *Unpredictability*—it is mostly unknown which are the "correct" or "expected" data values. This implies that the data optimization cannot be performed at the same time as the data collection. To predict the data quality, it is mostly necessary to analyze the entire dataset.

### 2.1. Big Data Reading and Storing

For an unknown big dataset, it is suitable to first consider the way of storing. That means to analyze data storing structures—the data representation format (textual or binary form), the data encryption type, etc.—clearly; one of the most important conditions prior to efficient data processing is the proper location for storing. The most frequent solutions today are XML files, distributed file systems [12], content delivery networks [13], clouds, or special databases [14]. The concrete efficient solution depends on the data type. Completely different solutions may be adopted, for example, for image datasets versus textual information from social networks.

### 2.2. Ecosystem and Tools for Big Data Processing

The preprocessing phase concerns data cleaning, outlier detection, normalization, interpolation of missing values, or noise filtering (noise reduction). For an unknown dataset being considered a high-quality dataset, it is practically impossible to perform these operations. The reason lies in the lack of knowledge of the data distribution before preprocessing, which makes it impossible to sort the data on the basis of criteria. The next steps of the processing phase concern dimensionality reduction, feature selection, and discretization. The general approach of selecting a representative subset of the data is not working very well here, because big datasets are mostly unstructured. A similar conclusion can be drawn for data normalization. There are several algorithmic approaches, designed especially for big datasets. Still, these approaches require prior knowledge relative to which the correct data values should be [15]. For the preprocessing of *big data*, distributed frameworks are used. The major platforms used are listed here below. Many of them are developed by the Apache Foundation in the framework of their Apache *big data* processing ecosystem.

(a) *MapReduce* [16] (Hadoop v2)—allows the storing of files via HDFS or the processing of stored data values. This solution uses an efficiently scalable distributed architecture.

(b) *Computing Engines* [17] (Spark, Storm)—approximately 100x faster than traditional MapReduce, used in many solutions.

(c) *Processing Pipelines* [18] (Kafka, Samza)—basically targeted on efficient caching of procedures that allow further processing.

(d) *Databases* [19] (Cassandra, Hive)—allow multiple-times faster data search in comparison to the traditional SQL-based approach.

(e) *AI-based frameworks* [19] (Mahout, ML over Spark)—help to use the traditional machine learning methods on *big data*sets.

### 2.3. The Problem of Big Data Quality Evaluation

The evaluation of the quality of an unknown big dataset is a highly complex and difficult task. For specific data, it is practically impossible to find missing values, check or normalize the values, and to detect the often-substantial amount of additional noise. Such datasets have in common that it is, in principle, unclear how to compare their values, or consider which values should be correct. Commonly known datasets can be evaluated by using the traditional approaches, which have been adapted to the scale of big datasets. The current analytic ecosystem solutions summarized above are able to process these datasets quickly and efficiently. However, due to the unstructured character of big data, it is not easy to reduce the data volumes to representative data sub-sets, which are much smaller, but preserve the original properties of the whole dataset. The identification of outliers represents a similar problem case, because it requires prior knowledge of predicted/predictable values. Data analysis and data mining cannot be considered separately from *big data* quality assessment [20], and it is necessary to use a large variety of these methods, or analytics, to discover whether valuable information or knowledge exists or not in *big data*, and whether this knowledge is useful or not. Poor data quality will lead to low data efficiency and produce decision errors. Given the data diversity, hierarchical structures for a data quality framework with assessment procedures [20] will not be readily adaptable to any kind of data but remain very specific, applicable only to a particular domain. Thus, the era of *big data* is inspiring the search for new approaches and algorithms. This search for new solutions will require exploring radically new possibilities in directions well off the already beaten tracks of computer science and incite data scientists to venture in territories nobody has ever ventured before.

### 3. *Big Data* Analytics: From Machine Learning to Artificial Intelligence

A major challenge for *big data* analytics consists of the accurate and fast segmentation of large datasets consisting of parameter values, generally expressed in numbers, into quality clusters, i.e., those which optimally define the input, which is less than a straightforward matter. For example, for a dataset with $n$ = 1000 feature rows, there would be $10^{301}$ possible combinations to evaluate in cases where prior knowledge relative to how the data are organized is unavailable. While the current trend is to believe that artificial intelligence and deep learning in increasingly larger neural network architectures will provide the all-encompassing answer to questions on how to address this problem, the scientific principle of parsimony (Occam's razor) implies starting by trying to improve the accuracy of data mining without "reinventing the wheel" altogether. This should be possible by building on machine learning algorithmic approaches which have proven their worth. When combined adequately in a building block approach, simpler approaches may prove far more powerful than other, more complex algorithms and improve the quality of data mining and software analysis, as will be illustrated later herein on the basis of specific examples.

Machine-learning (ML) systems [21] identify objects in images and select relevant results of search. The classic supervised machine learning approaches were initially limited in their ability to process data in its raw form and the analytic systems required specific domain expertise to program feature extractors that transform raw data into a valid internal representation vector for learning and input classification. This "old-fashioned" way of programming had the considerable advantage that the result obtained could be linked to well-identified and controlled steps in the machine-learning procedure. In the current context of *big data* analytics, unsupervised ML algorithms prevail. We propose that the most prevailing analytic approaches to *big data* may be arbitrarily ranked into three categories: (1) '*Educated*', (2) '*Wild*', and (3) '*Artificial Intellligence*' (AI); an overview is given in Figure 2.

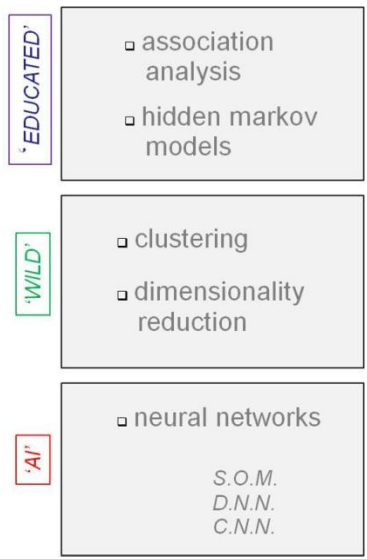

**Figure 2.** Analytic approaches to *big data*: From unsupervised machine learning ('educated' or 'wild') to artificial intelligence (AI).

Unsupervised ML methods of the 'educated' type (Figure 2, top) imply educated guesses (hypotheses) relative to the data structure and are aimed at generating predictions. The most common of these would be hidden Markov models, which have been applied in the *big data* context to generate spatio-temporal predictions based on clusters of interest in large video datasets [22], and association analysis or rule-mining, aimed at finding frequent items in a database with the least complexities to generate predictions. Association analysis for *big data* is limited by the problem of finding large enough item clusters in datasets [23]. Making assumptions about trends in large data to generate predictions makes no sense if nothing about the quality of the data is known a priori.

The unsupervised ML methods of the 'wild' type (Figure 2, middle) make no particular assumptions about the data, and are therefore a first choice for detecting quality and structure in unknown *big data*sets. These different methods can easily be adapted and combined for the identification of meaningful groups in large and unknown datasets. They employ algorithms for data clustering [24, 25] and dimensionality reduction, often by principal component analysis [24], to reduce the size of very large data in high dimensional (generally Euclidean) space to smaller sets of weighted data points, or core datasets. Essential properties of clustering algorithms as a first-choice analytic solution for detecting quality in *big data* will be pinpointed further in a subsection here below.

Neural network structures, which are inspired by functional properties of neurons in the primate brain, are synonymous with AI (Figure 2, bottom). The so-called deep neural networks (DNNs) [26] are the most recent and also the most complex breed of representation-learning algorithms, with multiple levels of representation obtained from groups of non-linear modules that transform representations at one level (starting with the raw input) into representations at higher, increasingly abstract levels. With many such transformations, increasingly complex functions can be learned. For classification tasks, higher layers of representation amplify certain input data (important for discrimination, and suppress irrelevant variations. The key aspect of deep learning is that it is, by nature, unsupervised, i.e., feature processing is no longer controlled by a human expert, and the models learn to process features from the initial input data fed into a given machine learning procedure, of which there are many [21]**.** Deep learning may outperform other methods in *big data* classification with respect to accuracy of prediction [27], but not necessarily in other specific tasks, such as image-based cell-type annotation, for example [28]. It is not always clear which processing step in a deep learning approach would account for better results obtained [29].

Convolution neural networks (CNN) are a type of deep learning models, often termed deep CNN [27]. They have an input layer, and output layer, and hidden layers. The hidden layers usually consist of convolution or pooling layers, and all layers are fully connected. Convolution layers apply a convolution operation to the input, then the information is passed to the next layer. Pooling combines the outputs of clusters of neurons into a single neuron in the next layer. Fully connected layers connect every neuron in one layer to every neuron in the next layer. In a convolution layer, neurons only receive input from a subarea of the previous layer. In a fully connected layer, each neuron receives input from *every* element of the previous layer. The network works by extracting features from images, which are learned while the network trains on a set of images. CNNs learn to detect features through tens or hundreds of hidden layers. Each layer increases the complexity of the learned features. The problem of any of these deep learning algorithms with respect to data quality detection is that they generally do not apply parsimony rules for eliminating excessively large amounts of random or meaningless data in a set [30]. This constitutes a drawback for any kind of data domain where precision of data description matters.

The simplest form of an AI neural network is the self-organized map (SOM), sometimes also called Kohonen map [31–33]. To provide a snapshot view of the difference in complexity between the SOM and a DNN, a graphic illustration of their functional properties is shown in Figure 3 here below. In its most elementary form, the SOM has a functional architecture with a single input layer, a single processing layer (neurons), and a single output layer. One of the advantages of the SOM analytic is that it reduces the input dimensionality in order to represent the input distribution as a map. In image analysis [34], such dimensionality reduction allows performing analyses of large image series with millions of pixels per image with to-the-single pixel precision. In its most elementary form, the SOM uses a winner-take-all learning strategy, and has hitherto unsuspected potential as a parsimonious and effective image data analytic, as will be explained in further detail in the subsection here below.

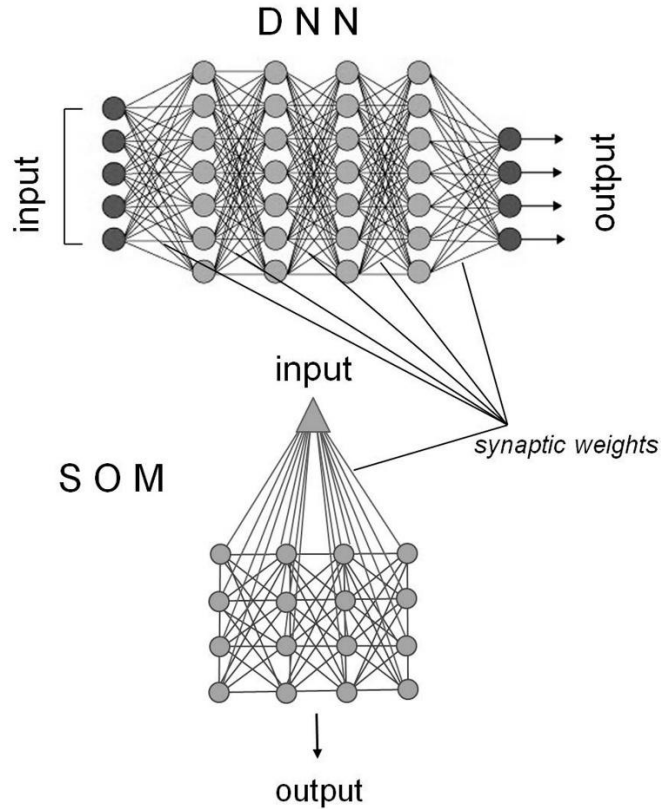

**Figure 3.** Functional architecture of a deep neural network (DNN) (top) with four fully connected hidden layers of neurons between input and output layers, and that of a self-organized map (SOM) with a single layer of fully connected neurons between input and output. The SOM reduces the input dimensionality in order to represent the input distribution as a map. For learning, it generally uses winner-take-all, while a DNN may use a multitude of learning algorithms, at multiple steps of processing.

### 3.1. Clustering Algorithms: Old Dogs for New Tricks

Clustering has been around for a rather long time [25] and, as has become clear from the overview given in the previous section, before diving into the deep blue sea of deep learning, clustering methods are among the most essential parsimonious analytics to consider first. They prove ideal candidates for building on simplicity in order to (1) improve the quality of large data analysis and (2) minimize the necessary computation times. Cluster analysis is well-suited for identifying meaningful groups in large and unknown datasets [35–47]. Algorithms such as "k-means" [37] use pair-wise similarity metrics such as the Euclidean to identify the cluster–vertex memberships, where the Euclidean is by far the most popular choice for "k-means" functions. Considerable reduction in computation times can be obtained through optimization of similarity metrics [47] and local heuristics [48].

For cluster algorithms to provide accurate results, characteristics in the training data need to be captured by the algorithm. Cluster algorithms are dedicated to specific data topologies, hence the need for data normalization. Yet, the smoothing of data results in a loss of information [49,53]. For a similarity metric to improve prediction quality, the similarity metric needs to reflect the core attributes of each dataset [48], a requirement that is satisfied the "hpLysis" software solution for example [53]. The use of data normalization strategies may hide key features of datasets [49], an observation that may explain, just to give an example, why genes with a strong differentiation are not always detected in microarray analysis [49]. The default strategy here consists of applying averaged normalization [53–58]. While [59] argues for tuning algorithms towards data topologies, this strategy is rarely applied. There is no "best clustering algorithm", and any algorithm imposes a structure on the data given [59].

To address issues relative to data topologies, [59–63], strategies for dimensionality reduction and feature extraction as a pre-processing step to clustering are proposed [61]. However, these strategies are known to provide a relatively poor performance [64] when compared with the fine-tuning of algorithmic building blocks [45]. Clustering algorithms are generally used for image segmentation, tracking of individuals in CCTVs, classification of digital signals, and so forth. Comparisons of measurements from 20+ cluster algorithms and 300+ correlation metrics have shown that established cluster algorithms have a speed and accuracy considerably higher than that of other recently proposed algorithms. Some work [49,50] identified 100+ new cluster algorithms that may be successfully applied to large datasets in fields such as digital signal processing and/or image analysis. Measurements across a large number of cluster algorithms and more than 100 real-life datasets reveal how the right parameter choice for clustering can outperform a large number of other recent algorithmic approaches to the problem of large datasets [65].

In the case of big data, the motivation is to increase the performance of clustering for data mining, and to identify new data mining algorithms without "re-inventing the wheel" altogether. This can be achieved, as argued here, by the integration of "trivial" algorithms to outperform more "complex" ones [48,49]. In this context, exploiting machine learning strategically to identify new clustering algorithms considerably improves the quality of image analysis and digital signal processing, for example. It represents a performance boosting strategy based on:

(1) Using building blocks from the parameterization of a large number of cluster algorithms to optimize the clustering accuracy,
(2) A combination of local and global heuristics that outperforms the speedups enabled by any single clustering algorithm, and
(3) An unsupervised automated (machine learning) procedure to construct many new clustering algorithms from the training data.

A major challenge when estimating a set of new algorithms consists of finding a precise definition that states, in clear terms, what exactly a clustering algorithm is supposed to do, at which level, and under which boundary conditions. In practice, there is no well-defined separation between algorithm, software, and software–algorithm configuration. Although mostly 'wild' by nature as clarified in Section 2 here above, clustering algorithms sometimes carry implicit (tacit) assumptions about input data and "ideal" cluster outcomes, i.e., heuristics, as is the case for column-based normalization, Kendall's Tau correlation metric [50,51,66], or the "Silhouette" convergence metric [67], for example.

In the context of *big data*, clustering algorithms have similar potential and face similar challenges as genetic algorithms (GA) and fuzzy clustering. This is due to the combinatorial space between "training data" versus "user hypothesis" versus "real-life data". For k-means, GA, and fuzzy clustering to provide accurate predictions, they may be adjusted to topological traits [68–77]. A genetic algorithm (GA) is understood as a "generic term subsuming all machine learning and optimization methods inspired by neo-Darwinian evolution theory" [74,75], and is often used in a combination with other methods for accurate inference of knowledge, like fuzzy clustering, which is a permutation of k-means [76]. There is no clear separation between GA versus k-means, and GAs may be merged directly with k-means [70–77], or k-means may be optimized through a GA seed selection [78–83]. In the context of big data, the use of classical clustering algorithms combined with accurate training is inseparable from the practical application of GA, or evolutionary algorithms in general. Dimensionality or model order reduction in very large data can be achieved by Fuzzy C-means clustering [84], or by astutely combining k-means with principal component analysis [85].

*3.2. SOM for Single-Pixel Change Detection in Large Sets of Image Data*

In the image and vision sciences, at the intersection between machine learning algorithms for computer vision and computational neuroscience, low-level artificial neural networks have been proposed well before the area of big data and deep learning neural networks. The earliest AI approach in this context is, as explained in Section 3 herein, the self-organizing map or SOM [31–33,86,87], is largely inspired by the functional properties of visual neurons in the non-human and human primate [88–91], and by psychophysical data on human detection [92–101]. How elementary output parameters of the SOM may be exploited for detecting certain qualities in large, arbitrarily archived image datasets may be illustrated here on the example of rapid, unsupervised detection of strictly local changes invisible to humans in image time series. Such local changes in images may reveal different meaningful states of a physiological structure, tissue, or cell, and reflect progression or recession of a pathology, or the progressive response of a cell structure to treatment, not detectable by any of the currently available medical image analysis tools.

A simple functional architecture of the self-organizing map may be applied to the unsupervised classification of massive amounts of patient data from different disease entities ranging from inflammation to cancer, as shown recently [35]. Other recent work [102–105] has shown the quantization error ($QE$) in the output of a basic self-organized neural network map ($SOM$); in short, the SOM-QE is a parsimonious and highly reliable measure of the smallest local change in contrast or color data in random-dot, medical, satellite, and other potentially large image data. The SOM is easily implemented, learns the pixel structure of any target image in about two seconds by unsupervised "winner-take-all" learning, and detects local changes in contrast or color in a series of subsequent input images with a to-the-single-pixel precision, in less than two seconds for a set of 20 images [106,107]. The QE in the SOM output permits to scale the spatial magnitude and the direction of change (+ or −) in local pixel contrast or color with a reliability that exceeds that of any human expert [102].

Applied to the automatic classification of arbitrarily archived scanning electron microscopy (SEM) images of CD4+ T-lymphocytes (so-called *helper cells*) with varying extent of HIV virion infection [101], a four-by-four SOM (for an illustration, see Figure 3) may be trained on any image of a given series, whatever its size, using unsupervised winner-take-all learning. The functional architecture of the SOM is minimal, with a constant number of neurons and a constant neighborhood radius [31–33,86,87]. Winner-take-all learning and subsequent SOM processing to generate QE output for the automatic classification of images from a series of 20 takes less than two seconds. Each of the images from the series is associated with a SOM-QE output value. These values are then automatically listed in ascending order of magnitude on the y-axis, the image corresponding to each QE value is ranked on the x-axis as function of its SOM-QE label. This analysis achieved a 100% correct classification of the CD4+ cell images in the order of the magnitude of HIV virion infection displayed in an image. A human expert electron microscopist with more than 25 years of experience took 52 min to sort the color-enhanced micrographs in the correct order, and 69 min to roughly sort the original grayscale images, but not with 100% correct. Moreover, for the human expert to be able to perform such a task at all, it is necessary to visualize all the images of a given series simultaneously on the computer screen, using software that allows zooming in and out of single images for close-up two-by-two comparisons. Thus, since quality and clinical meaning in larger sets of imaging data cannot easily be perceived by expert visual inspection, we need to resort to parsimonious and effective computational blocks to assist expert analysis as the image datasets grow larger. The SOM-QE outperforms the RGB mean in the detection of single-pixel changes in images with up to five million pixels [107], which could have important implications in the context of unsupervised image learning and computational building block approaches to large sets of image data, including deep learning blocks for automatic detection of contrast change at the nanoscale in transmission or scanning electron micrographs (TEM, SEM), or at the sub-pixel level in multispectral and hyper-spectral imaging data [108]. Whatever the method of image analysis, at the beginning, any input image is broken down into pixels, and the analytic process starts from there; from utmost simplicity towards increasing complexity. With respect to the principle of parsimony, the simplest network architecture that does the job should be the one that is chosen. However, this is not systematically adhered to in contemporary data science, where complexity often outvotes simplicity [109], and where data scientists can "learn" to build a DCNN "in just ten minutes" through *Google*. This is a hazardous trend [5,9], and we may lose the track of what works best, at which level of explanation, and within which limits. This may then result in seemingly successful data analysis by sheer "fluke", where nobody asks why exactly it works, and which processing step(s) in the model explain(s) the results obtained.

### 4. Can Cities Become Really "Smart" or Will the *Big Data Jungle* Continue to Proliferate?

Mark Weiser stated in his article "The computer for the 21st century" [110] back in 1991 that "the most profound technologies are those that disappear. They weave themselves into the fabric of everyday life until they are indistinguishable from it". Following this vision, Weiser is often seen as the forefather of ubiquitous computing [111], whereas he defined a smart environment as "the physical world that is richly and invisibly interwoven with sensors, actuators, displays, and computational elements, embedded seamlessly in the everyday objects of our lives, and connected through a continuous network". Thus, the evolution of connected devices of everyday life—commonly referred to as Internet of Things (IoT)—allows for an ever-increasing amount of data characterizing different facets of our everyday lives. When referring to "smart cities", a lot of different aspects have to be considered. As shown in Figure 4, inspired by [112], the landscape of building blocks of smart cities is very widespread. In addition to the illustrated areas where large amounts of unstructured data are generated (e.g., smart grid, smart health, smart people, smart and interconnected cars, etc.) a lot of other data-producing entities in different areas play a vital role in this ever-increasing amount of unstructured data. With the advent of inter-connected sensing devices delivering constant streams of data, the *big data* jungle is constantly growing and proliferating. Different applications exchange or produce information using embedded, virtual, or other sensor devices [113]. Be it a smart-home combined with smart-grid technology, or interconnected vehicles (i.e., "smart cars"), data-delivering smart phones, or even data extracted from online user behavior in the area of e-commerce [114–116]. Data are being generated everywhere at any time. Thus, with the technological capabilities and the sensing devices integrated into our daily lives, we have the capability to sense the world, and gather massive amounts of—rather unstructured—data. The problem is that we still massively lack in using this data wisely in order to (i) extract relevant information for specific tasks, (ii) recognize patterns and generate new information, or (iii) simply store and further process data. Possible reasons for this fact could be:

- The variety of different types of data and the common unstructured nature of data arising from (i) different application domains, (ii) different devices delivering the data, and (iii) proprietary solutions;
- The fact that there is not "the one big data" algorithm—artificial intelligence as a research field provides a lot of different algorithmic and technical approaches; nevertheless, all of them are application-specific;
- That data could be error-prone;
- Data could simply be misused (i.e., using data in a "wrong" way or within the wrong context or domain);
- Computing resources and tools are not braced for coping with massive amounts of data.

Taking into account all the aforementioned aspects, the research challenge in applied science has now shifted from data acquisition to data analysis. The data are there—or at least we believe that we know how to access them—and it is now time to extract relevance, to decide what matters, and what does not. The simple observation that the amount of data is steadily increasing has been subject to research in recent related work. For example, Marx [114] presents challenges for *big data* from the perspectives of medicine and biology and argues that a "data explosion" has happened (and is currently gaining in volume). Additionally, with such an increasing amount of data, the fact that in an entity on a city scale, where different areas (e.g., smart homes, smart grids, smart cars, connected people, etc.)—each for itself delivering massive amounts of unstructured data—deliver data, it is almost impossible to extract relevant information or detect quality and meaning [117]. Pruning the big data jungle is subject to research, whereas the recent advances in artificial intelligence seem to be promising approaches. Unsupervised learning [118] that intends to discover patterns in unlabeled and unstructured data or detect anomalies could be an algorithmic solution. As already shown and discussed in the article sections here above, there is not "the one algorithm for *big data* analysis". The complex problem space represented by *big data* cannot be addressed by any single, novel or established, analytic solution. The potential value of any computational solution will depend on the application domain, the characteristics of the dataset given, and the specific use case in its context. Thus, for each specific problem, the best algorithmic solution is bound to be different from that for other problems in the complex universe of *big data*.

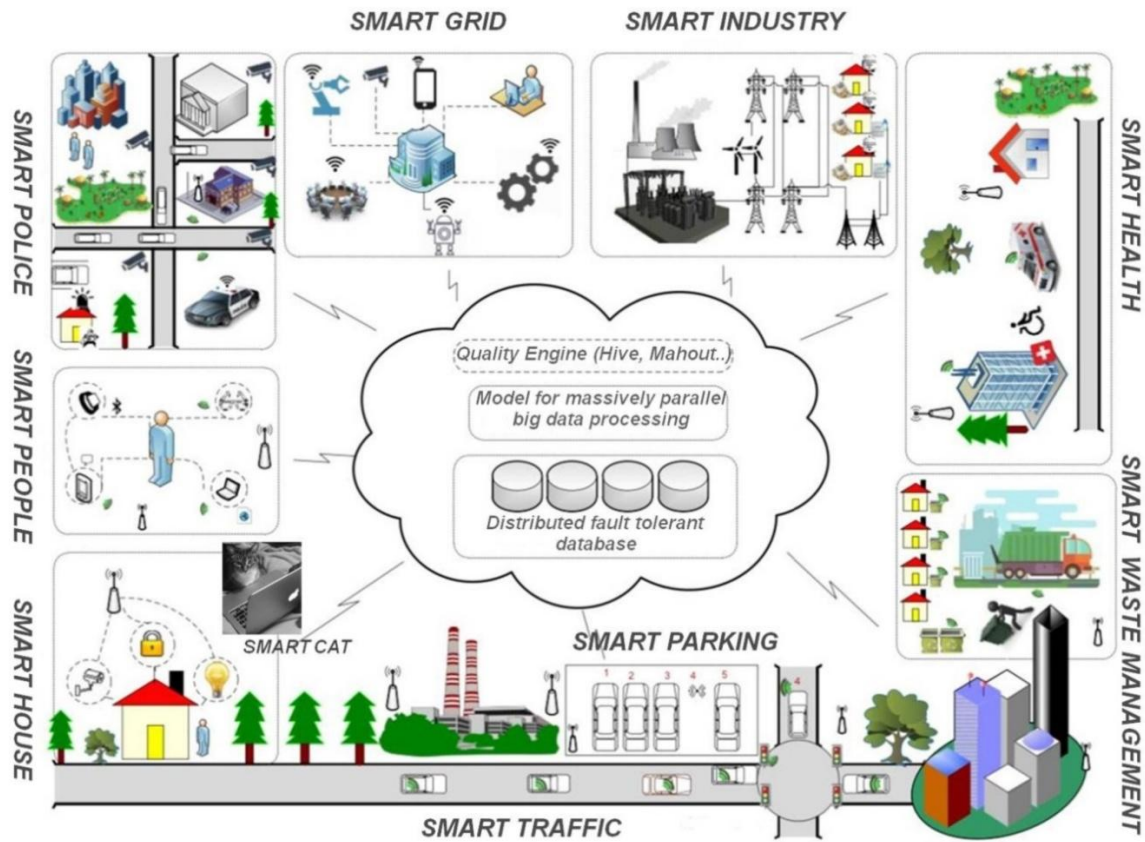

**Figure 4.** Idealistic view of a landscape of building blocks of *smart city* components through *big data* technologies, adapted from [112].

The sheer amount of data in itself poses a major problem, whatever the kind of data or application domain. As shown in Helbing et al.'s [117] model for digital growth (for an illustration see Figure 5), computational resources double about every 18 months (*Moore's Law* [119]) and data resources double about every 12 months. These two resources follow an exponential growth, while the processes to analyze data commonly follow a factorial growth resulting in large amounts of data that cannot be processed at all. Frangi et al. [120] refer to this phenomenon as "dark data"—i.e., the data that cannot be processed due to the high systemic complexity, information content, and the meaning they carry. The big data jungle is inexorably growing, since the world is polluted with sensors and data-delivering entities.

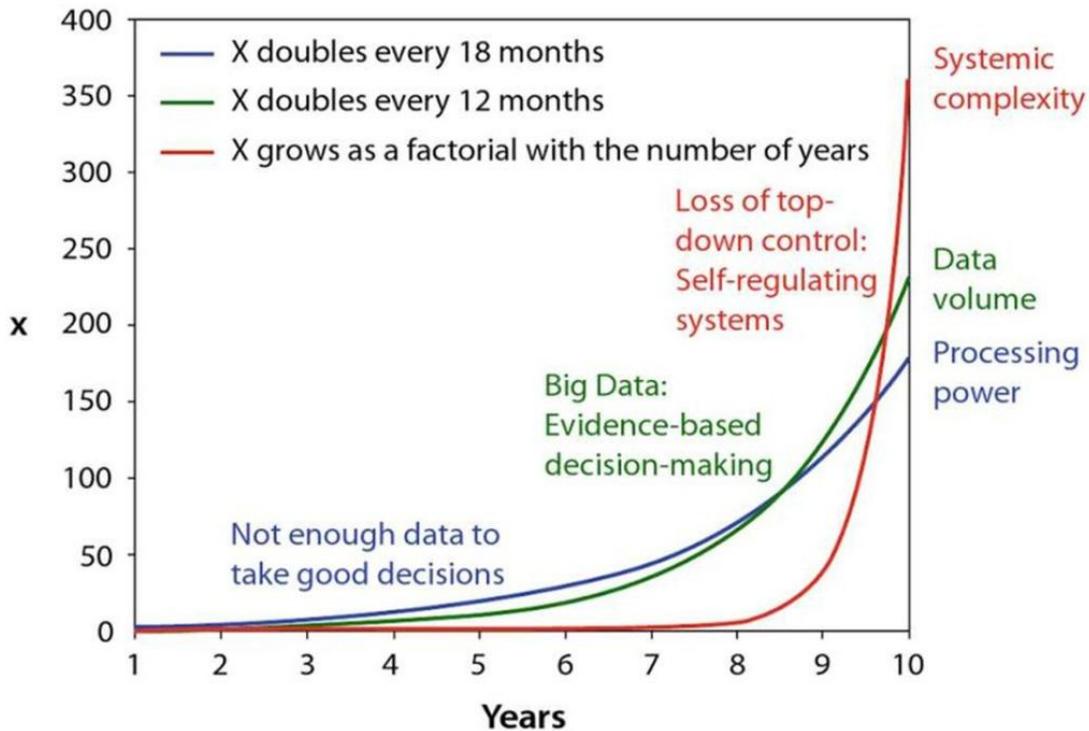

**Figure 5.** Helbing et al.'s [117] model for digital growth comparing computational and data resources with systemic complexity to analyze data.

## 5. The Chicken-or-the-Egg *Paradoxon* of Model Building or Why We Need Subjective Views and Pragmatic Methods to Analyze *Big Data* Contents

Data management for *big data* analysis requires particular attention, not only as a result of the quantity of data involved, but also because of the requirements of the data analysis process itself. Traditional database approaches concentrate on the *management* of data and of the schema behind it. More generally, we can, say, speak of *instances* and of a *model* for their description. A database model is predefined. It is used as a blueprint for newly created instances, it formulates rules and constraints by which to manage instances, and it defines how data can be queried.

One class of big data research studies the querying of large amounts of data, where queries require some model for the data. The kind of big data analysis addressed in this article works quite the other way around: Given data with "partially known" structures, find a model that generalizes properties of interest and that derives rules and constraints for datasets.

The ANSI-SPARC Architecture [121] that received particular attention for relational databases defines three levels of models: The *external* level that defines user-specific views on the data, the *conceptual* level that describes the whole data of a database logically, and the *internal* level that describes how the data are maintained on the physical level. Applying the architecture (in a kind of reverse manner) to big data analysis, we aim to find external views to describe particular aspects of a large set of data that has been created according to an (unbiased) conceptual model. Most information systems incorporating a database work with data that describe real-world entities. The data are a representation of things that exist outside of that system. This leads to the observation that a conceptual model serves two purposes at once: On the one hand, it is a technical description of the computational properties of data. For example, it determines that some data represent a non-negative integer number. On the other hand, it reflects an application domain. For example, it formulates that some data represent the weight of a person and thus that the unit of measure of grams is attached to it and that values typically lie in the range of 2000 to 200,000. While the former aspect of technical representation is mostly covered by type systems/database schemas, the latter aspect of domain-specific interpretation is buried in constraints and in application code that make use of the data.

Typically, a model is considered adequate if all data under consideration conform to it, if it allows capturing the meaning of a suitably large extent of the modeled application domain, and if it is minimal in the sense that no model element can be omitted without violating the first two requirements (*Occam's razor*). For all of these properties, it is required to know the bounds within which a model may be applied with respect to the application domain at hand. A first step towards deriving such a model is to find *content* in the set of data—that significant data that can be interpreted as a meaningful description of a domain entity [122]. There are two basic ways content can be represented by data: Firstly, the data can represent the entity directly, e.g., if those data are a (unique) product number that identifies some sales item, or if the data provide a photo of the entity. Secondly, the data can describe one property of the entity, and a sufficient number of such properties characterizes the entity.

Data representing entities directly serve as a sign as studied in semiotics. The triadic semiotics of Charles Sanders Peirce proposes a three-layered hierarchy of signifiers (*Firstness*, *Secondness*, and *Thirdness*) [123]. There are three kinds of representations [124]: On the first level, *icons* refer to entities directly. On the second level, an *index* has a relation to the signified entity. On the third level, a *symbol* denotes a rule that describes entities. This allows distinguishing data that directly signify an entity, data that establish an entity relationship, and data signifying a rule or constraint on entities. An instance on one layer is built upon ones on the next lower layer. Therefore, Peirce's semiotics indicates that data that directly signify entities are required as a basis for derived model elements like relationships and rules.

These different levels of signification are, e.g., similar to normal forms of relational database schema. Third normal form provides a kind of iconic signification by a primary key. Second normal form typically achieves attribute independence from partial key candidates by the creation of referenced tables, thus creating "Secondness" indexes. Tables in first normal form typically require recognizing structures in order to identify rows that together describe an entity. Therefore, such row sets are interpreted as symbols. The creation of hierarchies of entity descriptions through content has been studied in the *concept-oriented content management* (*CCM*) approach [125]. The most important insight is the subjectivity of the signs found. This is discussed below. Furthermore, it turned out that the ability to recognize signs depends on a conceptualization of the application domain. Dual to the signifying part of data, these findings directly relate to the descriptive parts of data that provide a conceptualization of entities. The simplest kind of such descriptive data is a reference to a concept (which in turn is represented by data, as a "Thirdness"). This is, e.g., employed on the Semantic Web [126].

In a model, a *concept* names a phenomenon in the domain, and rules and constraints may come with it. This way, concepts describe classes of domain entities. A technical concept is a data type ("integer") and might be given for a particular database. A domain concept captures domain knowledge ("typical human weight") and can, therefore, not be predefined in a generic way. It carries some domain semantics that may be given as *intensional semantics*, i.e., by rules and constraints that data of a certain class have to fulfill, or by *extensional semantics* through the set of data that it is assigned to [127].

Figure 6 illustrates this problem on the example of interpretations of the painting *Napoleon Crossing the Alps* by Jacques-Louis David (example taken from a CCM application [128]). The photo of the painting is an iconic representation of that painting and of Napoleon since he is depicted. To art historians, the image is also an index for Hannibal crossing the Alps since the scenery visually cites this historical event. The image is a symbol for equestrian statue, an object of art that is characterized by the ingredients to be seen in the picture (equestrian, horse, pose, …). The photo is intentionally defined by the concepts attached. The concept *Strength,* in turn, is extensionally defined by some further images. The two ways of defining concepts have been derived from the epistemology of Ernst Cassirer [129–131]. From our perspective, Cassirer's work contains a possible answer to the question whether to start with a model (as databases do) or to start with data (as data analytics do): Instances and models co-evolve.

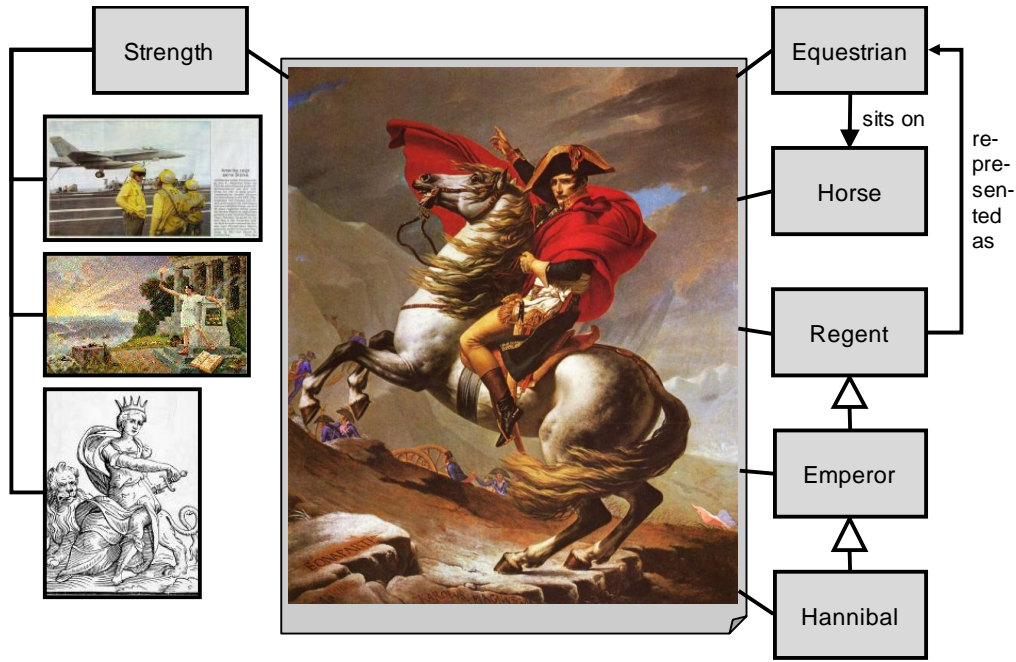

**Figure 6.** Intentional and extensional concept semantics in concept-oriented content management (CCM). Associations between images (Napoleon Crossing the Alps) and related concepts can be read in two ways. Sets of images define a concept, and concepts give meaning to an image.

Data analysis is performed for an *observer*, a user, or an algorithm that interprets the data. Such an observer will only perceive those data as meaningful for which there is a concept, and a concept will only be defined and used if there is sufficient evidence (in the form of data) for its existence. Which data actually constitute a sign depends on its observer, on the context the observer is in, and on the task at hand. An observer perceives the signs that are relevant with respect to those parameters. By the alternation of concept formulation and data interpretation, significant data are identified. It is used to build up hierarchies of signifiers to abstract from the concrete dataset and to carve out its content. The cycle of alternating concept creation and search for evidence for these concepts leads to relevance, and to the abovementioned urge to build minimal models. Only those concepts that are needed to describe instances and that cannot be expressed by other model elements will remain in the evolving model (*Occam's razor*). Overly complex concepts may be broken down into smaller sub-concepts if there is in turn enough evidence for these. Due to the dependency on observers' contexts and tasks, models are built in a pragmatic way. There will be no fixed syntax in the form of signifiers, and no fixed semantics given by concepts. Instead, the content found in a dataset will express a subjective view that is valid only for the context of the observer for whom it was defined.

Since both the identification of data that directly refer to a domain entity as well as the identification of data that describe (a set of) domain entities depend on an observer, there are coexisting views on the same data. These views need to capture a wider range of variants than the external views known for databases, though, since they do not only consist of restrictions of one global conceptual model, but instead offer freedom in the structuring of data and in the formulation of models. On top of that, models change over time. This is due to the co-evolution of instances and models, so that all newly identified data and every new model element can potentially lead to model additions and changes. Furthermore, users may want to adapt existing models to different, but related contexts in order to reuse results. Therefore, whichever method is used to analyze big data, the results in the form of significant data and models need to be stored in a personalized way for one particular observer. Personalization has become a common means in content management, but on a different level. The modeling requirements formulated here call for both model and content personalization capabilities.

In contrast to the need to formulate personalized models, users need to be able to communicate using data. Such communication takes place at least over time, because data are created and analyzed at different points in time. Furthermore, producers and consumers of data may differ. Communication requires shared models, though. Pragmatic communication in the face of personalization can be achieved by personalized models that are related to each other. Extensive personalization on the one hand and related models on the other have extensively been studied in the CCM approach. In CCM, model relationships are established by models derived from each other, and by reusing models for base domains in different personalized models.

The two dimensions of model relationships are illustrated by Figure 7. It shows four model constellations. In the lower left there is a singular model *Model₁*. To the right, a constellation in which *Model₁* incorporates (parts of) two other models: *Model₂* and *Model₃*. The graph in the upper left shows *Model₁* being refined by two personalized variants *Model₁₁* and *Model₁₂*. The edges depict the personalization relationships. The upper right of Figure 7 shows combinations of the two model relationships. Current research investigates the *minimalistic meta modeling language* (*M3L*) [132,133] that exhibits properties that make it suitable for the management of *big data* analytics results as proposed in this section. The M3L supports signification on the various levels. It regards context as a major modeling principle, and it allows contextual concept definitions. This enables different forms of personalization, variants, and model relationships between them.

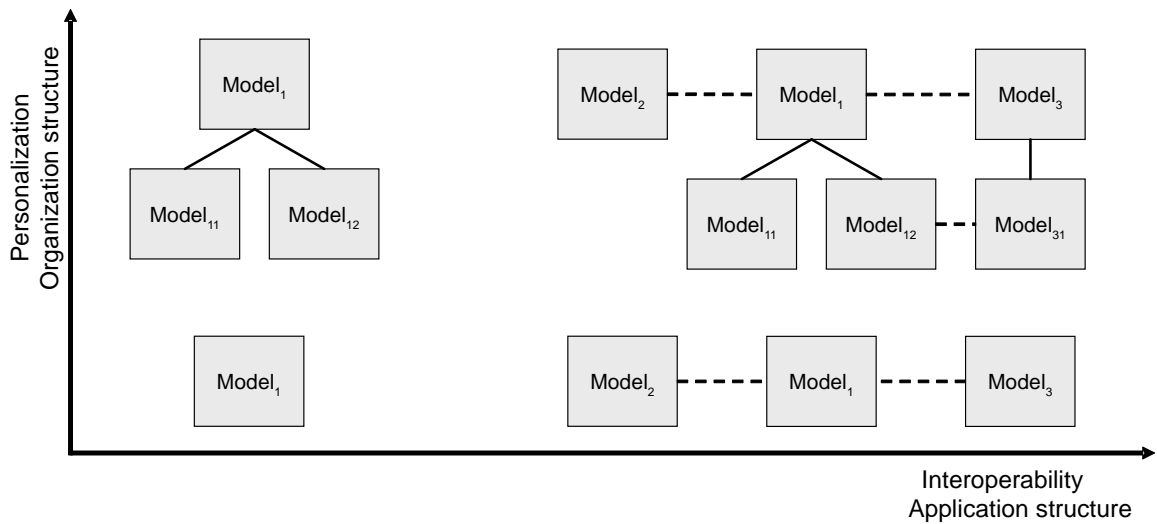

**Figure 7.** Dimensions of model relationships in CCM. Four model constellations that exploit model reuse and model personalization.

By its deductive properties based on contextual concept definitions and concept refinements, in conjunction with rules that allow both synthesizing concepts as well as matching instances, the M3L allows higher concepts that abstract from individual instances. Without introducing the M3L, we provide some examples to illustrate that. A concept like *Peter is a Person* is an icon for a person named *Peter* if the concept *Person* is given. Through a semantic rule that produces one concept from another, we can define indexical signification. For example, the definitions

*Person {Sex; Status}*

*MarriedFemalePerson is a Person {Female is the Sex; Married is the Status} |= Wife*

*MarriedMalePerson is a Person {Male is the Sex; Married is the Status} |= Husband*

establish an indexical relationship between *Married Female Person* and *Wife* (the former is directly resolved to the latter). The evaluation of M3L definitions allows deductions. For example, with the definitions

*MalePeter is a Peter {Male is the Sex}*

*MarriedPeter is a Peter {Married is the Status},*

the additional fact that

*MarriedMalePeter is a MalePeter, a MarriedPeter.*

results in the deduction that *Peter* is also a *Husband*.

Such M3L evaluations can be viewed as symbolic references. Each concept defines a context. All concepts in a context can be regarded as a model, so that contextual definitions allow establishing model relationships as, for example, in:

*A: Napoleon Crossing The Alps*

*A is a Model {NapoleonCrossingTheAlps is a HannibalPicture}*

*B is a Model {NapoleonCrossingTheAlps is an EquestrianStatue}*

*A2 is an A {NapoleonCrossingTheAlps is an ImageOfStrength}*

where the concept in *A2* is derived from the concept in *A* because the two models have a relationship (refinement, in this case, can be used for personalization). For model building, it is generally important to respect the boundaries of models with respect to the part of the application domain for which they are valid. The boundaries are required to judge which deductions can be applied to a dataset. It has to be noted that each deduction may lead to a pragmatic model change. Model boundaries are also relevant when transferring results across models along the relationships between these models.

To conclude, we can say that whatever means we use to analyze *big data*, it needs to be clarified how a given model is derived, which part of a domain it describes, and for what purpose. To this end, the means of management of the models and the information derived call for attention in addition to the investigation of methods for deriving models for *big data*sets. Such data management needs to be able to cope with model evolution as well as with coexisting models that allow different questions to be answered with the help of evidence contained in datasets.

## 6. Science and Culture from West to East or How We Relate Emotionally to *Big Data* and Artificial Intelligence

There has been an explosion in the amount of digital data since 2010 [134]. The amount of data is expected to reach 35 ZB in 2020. Over 95% of data were accumulated in this decade. Using this big amount of data has become practical owing to the advancement in digital technologies. We call it "*big data*" because it is a field that focuses on ways to analyze data, systematically extract information from them, or otherwise deal with <u>datasets</u> that are too large or complex to be dealt with by traditional <u>data-processing software</u> applications. However, this definition is fluid and can be understood in various ways. Thus, there is no clear definition for *big data*.

At the beginning of the first chapter of the famous book "BIG DATA", <u>Viktor Mayer-Schoenberger</u> and <u>Kenneth Cukier</u> [135] have taken up the Google Flu Trends (GFT). They studied how Google has mined a five-year log on the web, including hundreds of billions of searches on building an algorithm, and they claimed that "it is more effective than government statistics that cause delays in reporting", stating further that "we have built an influenza prediction model using 45 search words that have been proven to be a timely influenza index." Unfortunately, the GFT was not as effective as advertised. Shortly after the GFT was launched, the first problem occurred in 2009. It could not predict the swine flu epidemic at all. A report published in the Nature magazine in February 2013 observed that it predicted the influenza epidemic that occurred at the end of 2012 50% more severely than it actually was. Furthermore, the most inconvenient verification results since the launch of GFT were announced in March 2014. Media always emphasize the examples that work effectively. The apparent magic around "*big data*" is no exception here. However, there are many failures buttressing successful examples. Scientists have been continuously searching for theories of phenomena. Researchers tried to obtain a new theory that could explain the phenomena when no theory could explain it. These theories were based on causality. However, *big data* analysis does not have causality, and we have no golden tool for analyzing *big data*sets. There is no guarantee that a particular method of analysis that works successfully for one *big data*set will work for another. The reasons for this problem cannot be straightforwardly explained. Sometimes it works, and sometimes it does not and, therefore, dealing with *big data* is a big challenge.

Deep learning, very popular among data scientists, faces the same challenge. The problem is clearer in the case of deep learning than in that of *big data*. Although deep learning algorithms have many parameters, the principle of deep learning is always the same. Deep learning creates deep learning systems in minimizing cost functions by using a large amount of learning data. However, we do not know how to adjust such a complex system when it does not work well. The same challenge arises with *big data*, and the result is what matters most. The cause is largely unknown, but *big data* and deep learning may perform quite well, sometimes. We cannot discover the solution to cases where things do not work efficiently. Shall we call these new technologies science, or engineering? They are, in fact, a black box. Can we entrust our future to a black box? It is good when situations are convenient, but convenience often causes us to lose certain abilities. We forget how to read maps when GPS is practically applied everywhere. We have learnt how to calculate the square root a few decades ago, but presently, few people will be asked to give the square root with paper and pen, since anyone can easily find the square root of any number using a smart phone.

*Big data* and deep learning are tightly linked together [136,137]. As a consequence, this connection will be advanced further in various fields of study. When you fall sick in the future, the doctor might say "let us ask the black-box about your symptoms." How are we to feel about this? So-called technological singularity is an unavoidable issue when we discuss *big data* and deep learning [138]. The abilities of artificial intelligence (Al) may well exceed that of human beings by the year 2045. There is a high risk that about 47% of all U.S. jobs will be automated in the next 10 to 20 years [139]. Bill Gates argued that AI is a threat to humanity shortly after Microsoft's top announces that Al is not a threat to humanity. In addition, Stephen Hawking considered Al dangerous. However, there are different opinions regarding the current Al technology. AI could not pass the university entrance exam, even though it could win in a quiz show [140]. Deep learning, as the most common type of AI technology, is presently not real learning. However, having the name learning is somewhat superficial. The current technology cannot create intelligence beyond human capabilities. If that is the case, why should the singularity that AI surpasses human intelligence be discussed? Furthermore, why should technological singularity, where AI surpasses human intelligence, be discussed at all? We do not hear that technological singularity is considered a problem among experts in Japan. It is only the media and the non-specialist community of people that worry about technological singularity. Beginning with Kurzweil, who advocated it, people in science and technology who are the keenest on discussing technological singularity as a challenge for humanity are, indeed, mostly Westerners.

In our modern times, we have accumulated a vast amount of scientific and technical knowledge, but our education has been strongly affected by our different histories, cultures, and religions. Western culture and thinking have been influenced mainly by Christianity, a monotheistic religion. In contrast, the Japanese religion, Shinto, is polytheistic like the religions of ancient Rome and Greece. Of course, we accept and understand religious differences. In 2016, the BBC compiled a list of the 21st Century's 100 greatest films [141], which included "Spirited Away," a Japanese animated film by Hayao Miyazaki. The film was made for young teenage girls. In the film, a girl's parents accidentally crashed a party held by the gods, who changed them into pigs. Consequently, they lost their memories and lived just like pigs. However, their daughter was able to restore them to their human forms. Such a story is very difficult for monotheists to understand, but the BBC chose "Spirited Away" as number four on their list of greatest films ever. This choice is a reflection of current popular understanding of religious differences, but individual thinking is still strongly influenced by our individual religions.

Back in the 19th century, science advanced dramatically and people believed that humans would be able to create other human beings in the future. The novel *Frankenstein* is an excellent example for illustrating this [142]. Although *Frankenstein* has been made famous through films, the original novel is quite different from the films, where Frankenstein's creation, an artificial but biological man, is depicted as very large, powerful and violent while having rather low intelligence. In contrast, the novel depicts him as intelligent enough to master multiple languages. However, he is rather ugly and disliked by people because of his looks. At the end of the novel, Dr. Frankenstein deservedly dies for having created an artificial human from body parts of dead people, implicitly considered a sin in Christianity, where only God is to create humans and where the dead are to be left in peace and not to be exploited for unholy purpose. When Westerners discuss attempts at creating other artificial forms of life, bearing in mind that all life is sacred in Western religions, they may indeed believe, consciously or unconsciously, that any such attempt is "sinful". Yet, such belief can be altered, i.e., effectively manipulated. Consider, for example, the visual appearance given to current products of "popular" artificial intelligence.

Figure 8 here above shows a cartoon character (on the left) named "Astro Boy" [143], who is a super robot that can fly, possesses the strength of one million horsepower, and understands all languages. Although "he" is an android, Westerners tend to think of "him" as *cute, cool*. Not many Westerners are likely to think of him as *eerie*, *scary*, or *weird*. In contrast, what do we feel when we see the female humanoid robot [144,145] in Figure 8 (on the right)? "She" too is an android, "her" name is Madoka Mirai, and "she" probably looks uncanny to many Westerners. Although both creations are androids, Astro Boy is mostly considered *cute*, whereas Madoka Mirai may be considered *eerie* or *scary* by most in the Western culture. What makes the difference between these two? An important one is almost certainly Madoka Mirai's close resemblance to a human being. Just as the creation of another human was considered a sin in Mary Shelley's *Frankenstein*, people in Western cultures may feel similarly when they see this humanoid robot. Eastern Asians, such as the Japanese, Chinese, and Koreans, do not feel upset about artificial life and humanoids at all. There is most certainly a close connection between religion, collective belief systems, and how people from different cultures feel about artificial intelligence (AI) and its products. Deep learning technology for *big data* mining inevitably involves AI. Currently, it is impossible to use this technology for creating new forms of AI that would surpass human intelligence. AI still can only respond to questions that already have their answers, but it cannot provide solutions to unanswered questions that we face in data science every day. Furthermore, possessing the ability to solve problems is completely different from possessing willpower, motivation to act, and, ultimately, consciousness. At present, AI cannot understand our culture. In the year 2045, how will AI think and feel about the androids shown in Figure 8?

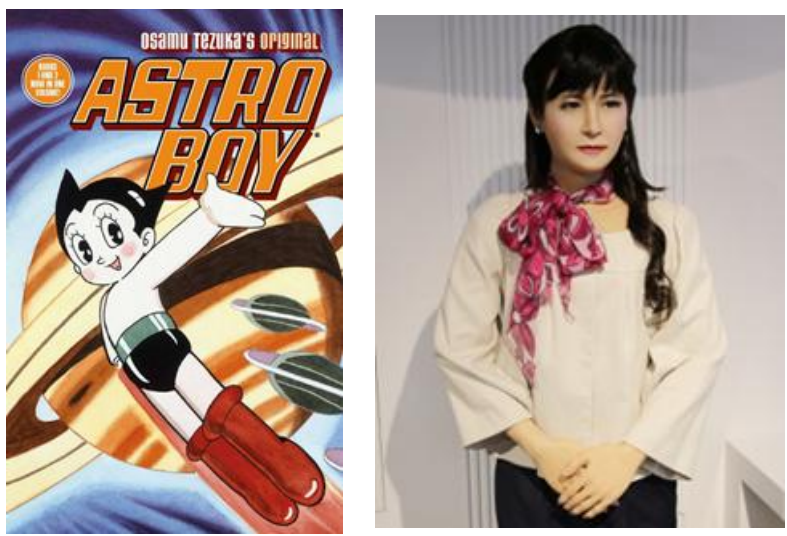

**Figure 8.** The robot Astro Boy (**left**) and the humanoid Madoka Mirai (**right**).

## 7. Conclusions

We still have limited understanding of how it will be possible to translate the *big data* potential into actual scientific, social, and economic value. The future trend in data science is announced as one towards fully automated data analytics, which implies artificial intelligence-based analysis and decision making in all fields [146], in other words: AI everywhere https://www.cioreview.com/news/artificial-intelligence-is-everywhere-nid-28010-cid-175.html.

This trend has raised high hopes in different fields of science for a better and faster understanding of structural complexity in living system [147–149]. The key argument defended in this review article here is not one against AI, but against any systematic replacement of the classical hypothesis-driven methods by new, increasingly complex and therefore difficult to control, data analytics that have not yet proved their worth. The different domain examples discussed in this review were aimed at illustrating some of the reasons why the generation of novel, data-driven hypotheses needs to be based on interpretable models. This will always and inevitably require validation and experimental testing. Complexity in analytical approaches does not necessarily outperform simplicity [150]. The currently emerging counter trend against fully automated *big data* analysis is that of exploratory data analysis http://www.creative-wisdom.com/teaching/551/Reading_materials/Yu_EDA_Oxford.pdf largely inspired by the philosophical and analytical work of John W. Tukey [151], who compared valid data (*big* or small) analysis to good detective work. This work should not be fully automated but may and will benefit from wisely and adequately developed artificial intelligence. Good detective work in data science has to

- Start by asking the right question(s);
- Look for the right clues;
- Draw the right conclusions from the clues available.

We are, thus, led to reconsider the somewhat inflexible principle of parsimony (*Occam's razor*) in traditional science within the context of contemporary data science, under a new and different light, where the advantages of simplicity have to be weighed against a seemingly growing need for complexity. This new context ultimately requires responsible decision making by domain experts, on a day-to-day basis.

*Big data* have not appeared all of a sudden out of nowhere but are a result of our cultural and technological development. They may be the ultimate consequence of the evolution of our species that will lead directly to our own extinction [152], programmed by ourselves, and taking final shape in fully autonomous AI. Since we may not wish to be kept as pets by fully autonomous monsters (they may not even want to keep us as pets!), we may finally accept that the time for our extinction has come, with the planet's resources getting scarcer. Soon there may not be enough food and water for everyone everywhere on Earth. Fully autonomous AI does not need water, air, or food—we do. However, there is also hope. Fully autonomous AI will consume a lot of energy and we may realize that we should save it for better purpose. Also, modern data science is not evolving behind closed doors, or in a void. Analytic systems aimed at capturing what is deemed to make sense in all these unstructured data are designed by domain experts, and the procedures and algorithms these analytics use are based on scientific reasoning. They can be, and generally are, tested and refined through scientific investigation. They can also be reassessed in the light of progress. Inductive strategies for identifying patterns within data do not occur in a scientific vacuum, they are discursively framed by previous findings, theories, and speculations or intuitions grounded in experience and knowledge. The fear of a data science of the future, where insights and knowledge are automatically generated without asking the right questions, is therefore no more, but also no less, than a collective fantasy. As we have all learnt from history, collective fantasies can be dangerous. Yet, new analytics and algorithms most certainly arise scientifically, not arbitrarily, to be tested in the light of state-of-the-art expert domain knowledge. Ultimately, we may hope that this will be enough to ensure that only the methods and algorithms that have proven their true worth will survive the *big data* revolution. The latter was brought upon us directly by *our* cultural and technological development. Without it, there would be no need for paradigm shifts in science, or highly developed artificial intelligence. At present, science is still struggling to find clear general definitions and guidelines for key concepts directly related to *big data* ("data science", "deep learning", "artificial intelligence"), yet, these concepts are presumed to enable us to cope with the problem represented by *big data*. This somehow seems to require leaving the beaten tracks of science. The logic of scientific explanation according to William of Occam's *summa logicae* [4] requires that the nature of the *explanandum*, or what is to be explained, is adequately derived from the *explanans*, or explanation given. Considering the case of *big data*, these are no more than a particular expression of any current *explanandum*; *big data* neither go along consistently, nor systematically (i.e., predictably) with any current approach exploited to model them. If data science does not tread off the beaten tracks of traditional science very carefully, it may end up not seeing the forest for the trees in the *big data* jungle.

**Funding:** This research received no external funding.

**Acknowledgments:** We thank our colleagues from our research teams for their support.

**Conflicts of Interest:** The authors declare no conflict of interest.

e, reference material.